\documentclass[prd,twocolumn]{revtex4}
\usepackage{bm}
\usepackage{epsfig}
\usepackage{amsmath}
\newcommand{\be}{\begin{equation}}
\newcommand{\ee}{\end{equation}}
\newcommand{\ba}{\begin{eqnarray}}
\newcommand{\ea}{\end{eqnarray}}
\newcommand{\bml}{\begin{mathletters}}
\newcommand{\eml}{\end{mathletters}}
\newcommand{\bes}{\begin{subequations}}
\newcommand{\ees}{\end{subequations}}

\newcommand{\bi}{\begin{itemize}}
\newcommand{\ei}{\end{itemize}}

\begin{document}
\title{A model of electroweak-scale right-handed neutrino mass}
\author{P.Q. Hung}
\email[]{pqh@virginia.edu}
\affiliation{Dept. of Physics, University of Virginia, \\
382 McCormick Road, P. O. Box 400714, Charlottesville, Virginia 22904-4714,
USA}

\date{\today}
\begin{abstract}
If neutrino masses are realized through the see-saw mechanism, can the 
right-handed neutrinos be produced and detected at present and future
colliders? The answer is negative in the most popular see-saw scenarios
for the simple reason that they are too heavy in these models.
However, a simple extension of the Standard Model (SM) particle content,
including mirror fermions, two $SU(2)_L$ triplet  and one singlet Higgs 
fields, leads to
a scenario in which the see-saw mechanism is realized with
the Majorana mass $M_R$ of the right-handed neutrino 
being of the order of the electroweak scale or smaller.
A custodial $SU(2)$ symmetry arising from the two triplet Higgs fields ensures
that $\rho=1$ at tree level even when their vacuum expectation values (VEV)
which determine the value of $M_R$, can be as large as the electroweak scale.
$M_R$ is found to obey the bound
$\frac{M_Z}{2} \leq M_R < 246\, GeV $ which makes it
accessible experimentally (Tevatron, LHC or ILC) since, in our scenario,
$\nu_R$'s can couple directly to the Standard Model (SM) gauge bosons.

\end{abstract}
\maketitle

%\section{Introduction}

Two of the most important experimental discoveries in the last decade
are undoubtedly neutrino oscillations and the accelerating universe.
Although the knowledge of individual neutrino masses is yet to be determined
experimentally, the most plausible explanation for the oscillation
data is the assumption that neutrinos have masses. When combined with
cosmological constraints, the picture that emerges is one in which
those masses are {\em tiny} of $O(<1\,eV)$ \cite{WMAP3}. 
However, its nature- Dirac
or Majorana- is unknown at the present time. 
%This crucial question
%will presumably be resolved if future experiments such as the neutrinoless
%double beta decay experiment gives a clear positive or negative signal.

By far, the most popular scenario for neutrino masses is the celebrated
see-saw mechanism \cite{seesaw} where neutrinos are of Majorana types.
It is well-known that in this class of scenario, small neutrino masses
arise because of a large hierarchy between between a Dirac mass
$m_D$ (typically of the order of a charged lepton mass) which
is intrinsically linked to the electroweak scale and a Majorana mass 
$M_R \gg m_D$, in the form $m_D^2/M_R$, where
$M_R$, the Majorana mass of the 
``right-handed'' neutrino, is typically some Grand Unified (GUT) 
mass scale or at least several
orders of magnitude larger than the electroweak scale. The high value
of $M_R$ makes this sector inaccessible experimentally.
One has to resort to indirect methods such as neutrinoless double
beta decays to probe the Majorana nature of the {\em light} neutrino.
If $M_R$ can be found to be of the order of the electroweak scale, one
could directly look for its signatures at future colliders through
the production and detection of right-handed neutrinos. This possibility
is realized in a model presented below.  

It is not unreasonable
to ask the following question: Could one obtain the see-saw mechanism
strictly within the SM $SU(3)_c \otimes SU(2)_L \otimes U(1)_Y$
by just extending its particle content? If it is possible to do
so, what would be the constraints on the Dirac and Majorana mass scales? 
What would the theoretical and 
experimental consequences be other than  providing
a model for the size of the neutrino masses? It is argued in this paper that
such a model can be constructed, with interesting
implications: $M_R$ cannot be larger than $\Lambda_{EW} \sim 246\,GeV$ and
the Dirac mass is unrelated to $\Lambda_{EW}$. What is the most
economical way to accomplish this?

The presentation will be organized as follows. 
It will be shown below that one can construct 
the see-saw mechanism by just staying
within the SM {\em gauge group} and by simply extending the SM particle content
to include mirror fermions, two additional triplet and one singlet Higgs fields.
(It should be emphasized that what we mean by mirror fermions are
simply fermions that behave like the SM ones but have opposite chiralities.)
$M_R$ is found to come from the VEVs of the triplet Higgs fields which can be
as large as the electroweak scale because of the existence of a custodial
$SU(2)$ symmetry which ensures that $\rho=1$. The paper ends with
a brief discussion of the possible signatures of our scenario such as
the production of the right-handed neutrinos through the process
$q + \bar{q} \rightarrow Z \rightarrow \nu_R +\nu_R$ with 
$\nu_R$ decays being characteristic of their Majorana nature. As we
will show at the end of the paper, the characteristic signatures are
like-sign dilepton events: a high energy equivalent of the
neutrinoless double beta decay.

Although the model presented here stands on its own, it can be seen to
fit into a grand unified model based on the
group $E_6$ which is designed to embed the SM and a new unbroken
gauge group $SU(2)_Z$ \cite{hung1}  as described in \cite{hung2} which
%This particular unification 
necessitates the 
introduction of the aforementioned heavy mirror
fermions \cite{roos}. At this point, it is worthwhile to emphasize
the fact that the main results of this paper rest on the assumption
that right-handed neutrinos belong to $SU(2)_L$ doublets but not
on the details of the so-called mirror fermions. In particular, the
possible observation of the ``light'' right-handed neutrinos does not
depend on the existence of heavy mirror quarks for example, although
anomaly cancellation will require their existence despite the fact that
they play no role in the subsequent discussion.

Let us first start out with the following SM particle content:
$l_L$ and $\Phi$, which are respectively the lepton
doublet and the SM Higgs field.
In addition, one has $e_R$ which is $SU(2)_L$ singlet.
The product $\bar{l}_{L} \Phi$ or
$\bar{l}_{L} \tilde{\Phi}$ contains an $SU(2)_L$ singlet. Since $e_R$
and $\nu_R$ are both $SU(2)_L$ singlets, the respective Yukawa couplings
to the above products are allowed giving rise to the ``normal'' Dirac
mass terms. If, however, $\nu_R$ is {\em not} a singlet of $SU(2)_L$ (nor
a triplet for that matter), it follows that $\bar{l}_{L} \tilde{\Phi}$
cannot couple to $\nu_R$. For definiteness, let us take $\nu_R$ as belonging
to an $SU(2)_L$ right-handed doublet -the so-called mirror leptons- as follows:
\be
\label{rhdoublet}
l^{M}_{R} = \left( \begin{array}{c}
\nu_R \\
e^{M}_{R}
\end{array} \right) \,,
\ee
where the superscript $M$ stands for {\em mirror fermions}. 
Just as with $e_R$ which is $SU(2)_L$ singlet, let us
also assume the existence of a charged {\em left-handed} $SU(2)_L$ singlet
mirror lepton, $e^{M}_{L}$ (the mirror counterpart of $e_R$).
Notice that, in this case, anomaly cancellation operates entirely
within the ``lepton'' sector (normal and mirror) as well as within
the ``quark'' sector.
A bilinear such as
$\bar{l}_{L}\, l^{M}_{R} = \bar{\nu}_L \nu_R + \bar{e}_L \, e^{M}_{R}$ 
transforms either as a singlet or as a triplet of $SU(2)_L$. Let us for
the moment assume the existence of a {\em singlet} scalar field $\phi_S$ which
can couple to that fermion bilinear. We have
\ba
\label{singlet}
{\cal L}_S &=& g_{Sl} \, \bar{l}_{L}\, \phi_S \, l^{M}_{R} + H.c. 
\nonumber \\
&=& g_{Sl}\,(\bar{\nu}_L \nu_R + \bar{e}_L \, e^{M}_{R})\,\phi_S
+ H.c. 
\ea 
%With this, one can write the following
The SM Yukawa couplings are given by
\bes
\be
\label{yuk1}
{\cal L}_{Y1}= g_{l}\, \bar{l}_{L} \, \Phi\, e_R + H.c. \,,
\ee
\be
\label{yuk2}
{\cal L}_{Y2}= g^{M}_{l}\, \bar{l}^{M}_{R}\, \Phi \, e^{M}_L + H.c. \,.
\ee
\ees
With the following VEV's:
\bes
\be
\label{vevs}
\langle \phi_S \rangle = v_S \,,
\ee
\be
\label{vevsm}
\langle \Phi \rangle = (0, v_2/\sqrt{2}) \,,
\ee
\ees
one obtains from Eqs. (\ref{singlet}, \ref{yuk1}, \ref{yuk2}) the 
following masses and matrices
\be
\label{neumass}
m_{\nu}^D = g_{Sl} \, v_S \,,
\ee
for the Dirac neutrino and
\be
\label{charged}
M_l = \left( \begin{array}{cc}
m_l& m_{\nu}^D \\
m_{\nu}^D & m_{l^M}
\end{array} \right) \,,
\ee
for the charged SM and mirror leptons.
The Dirac neutrino mass in (\ref{neumass}) is obtained from 
(\ref{singlet}) and, in (\ref{charged}), $m_l = g_l\, v_2 /\sqrt{2}$,
$m_{l^M} = g^{M}_{l} v_2 /\sqrt{2}$, with the off-diagonal mixing
being identical to the Dirac neutrino mass from (\ref{singlet}). The
diagonalization of (\ref{charged}) gives the following eigenvalues
for the charged lepton and its mirror counterpart
\bes
\be
\label{ml}
\tilde{m}_l = m_l\, - \, \frac{(m_{\nu}^{D})^2}{m_{l^M} - m_l}  
\ee
\be
\label{mlm}
\tilde{m}_{l^M} = m_{l^M} \, + \, \frac{(m_{\nu}^{D})^2}{m_{l^M} - m_l} \,.
\ee
\ees
We will assume that $m_{l^M} \gg m_l$. Furthermore, it will be
seen below that 
$m_{\nu}^D \ll m_{l^M},\,m_l$ and one can easily see that the mass mixing
in (\ref{charged}) is {\em negligible} giving $\tilde{m}_l \approx m_l$
and $\tilde{m}_{l^M} \approx m_{l^M}$. 

Before turning to the subject
of this paper which is neutrino masses, we wish to mention that
the coupling to $\phi_S$ which mixes SM fermions to their mirror
counterparts, also applies to the quark sector. 
Although the results of this paper do not depend on the existence of
mirror quarks, anomaly cancellation requires it. The
following short paragraph is written simply for the purpose
of completeness. It is straightforward
to generalize Eq. (\ref{charged}) to include the quarks. First, let
us denote the quarks by $q_L = (u_L, d_L)$, $u_R$, $d_R$, 
and their mirror counterparts by $q^{M}_{R} = (u^{M}_R, d^{M}_R)$, 
$u^{M}_L$, $d^{M}_L$. Replacing $g_{Sl}$ by $g_{Sq}$ in 
Eq. (\ref{singlet}) for the quark sector and noticing that
$m_q = g_q\, v_2 /\sqrt{2}$ and $m_{q^M} = g^{M}_{q} v_2 /\sqrt{2}$,
$M_q$ is obtained from (\ref{charged}) by the replacements:
$m_l \rightarrow m_q$, $ m_{l^M} \rightarrow m_{q^M}$, and
$m_{\nu} \rightarrow m_{\nu} (g_{Sq}/g_{Sl})$. The eigenvalues
are now
$\tilde{m}_q = m_q\, - \, \frac{(m_{\nu}^{D})^2\,(g_{Sq}/g_{Sl})^2}{m_{q^M} - m_q}$,
$\tilde{m}_{q^M} = m_{q^M} \, + \, 
\frac{(m_{\nu}^{D})^2\,(g_{Sq}/g_{Sl})^2}{m_{q^M} - m_q}$.
(We will assume $m_{q^M} > m_q$.) Again, one has $\tilde{m}_q \approx m_q$,
$\tilde{m}_{q^M} \approx m_{q^M}$. One can straightforwardly
generalize the above discussions to three families. 

The above exercise shows that the neutrino Dirac mass can be {\em independent}
of the electroweak scale in this simple ``model''. If this were
the whole story, one could simply make the Dirac mass naturally
small by having a small $v_S$ which is not constrained by any
other considerations. However, the total width of
the Z boson rules out that option because the addition of
the right-handed neutrinos to the coupling would increase the
neutrino contribution by a factor of 2. Therefore, in our scenario,
the right-handed neutrinos have to be {\em heavier} that half the Z mass.
This statement is rather general in the sense that, if $\nu_R$
transforms {\em non trivially} under $SU(2)_L \otimes U(1)_Y$, 
a small pure neutrino Dirac mass term is {\em forbidden}. (This
is not the case if $\nu_R$ is $SU(2)_L \otimes U(1)_Y$ singlet
in which case one should fine tune the Yukawa coupling, i.e.
$g_{\nu} \sim 10^{-11}$, in order for $m_{\nu} \sim
O(\leq eV)$.)

Since $l_R^{M,T} \,\sigma_2\,l_R^{M}$ (fermion bilinear
for $\nu_R$ Majorana mass term)
transforms as $(1+3,Y/2=-1)$ under $SU(2)_L \otimes U(1)_Y$, the
appropriate Higgs field cannot be a singlet which carries a charge
$+1$ since its VEV would break charge conservation. This leaves
us with the option of a triplet Higgs field $\tilde{\chi} = (3, Y/2=+1)$.
Explicitly, $\tilde{\chi}$ is given as
\be
\label{delta}
\tilde{\chi} = \frac{1}{\sqrt{2}}\,\vec{\tau}.\vec{\chi}=
\left( \begin{array}{cc}
\frac{1}{\sqrt{2}}\,\chi^{+} & \chi^{++} \\
\chi^{0} & -\frac{1}{\sqrt{2}}\,\chi^{+}
\end{array} \right) \,.
\ee
One can have a gauge invariant Yukawa coupling of the form
\be
\label{majorana}
{\cal L}_M = g_M \,l^{M,T}_{R}\, \sigma_2 \,\tau_2 \,\tilde{\chi}\, l^{M}_{R}\,.
\ee
With 
\be
\label{delta0}
\langle \chi^{0} \rangle = v_M \,,
\ee
one can see from (\ref{majorana}) that the right-handed neutrino acquires
a Majorana mass $M_R$ given by
\be
\label{majmass}
M_R = g_M\,v_M \,.
\ee
Notice that the use of an $SU(2)_L$ triplet had been made before
in the context of the left-right symmetric model \cite{LR} and
is very different from the present model. Here it is responsible for the
Majorana mass of the {\em right-handed} neutrino.

Without further restrictions, one could also have a Yukawa coupling of the type
$g_{L}\, l^{T}_{L}\, \sigma_2 \,\tau_2 \,\tilde{\chi}\, l_{L}$ 
which would give
a Majorana mass $g_{L}\,v_M$ to the left-handed neutrino. Unless $g_{L}$
is unnaturally fine-tuned to be very small, the presence of this term would
destroy the motivation for the see-saw mechanism. To forbid its presence
at tree level, one
could impose the following global symmetry $U(1)_M$ under which we have
\be
\label{U1M}
l^{M}_{R}, e^{M}_{L} \rightarrow e^{i\,\theta_M} l^{M}_{R}, e^{M}_{L} \,,
\tilde{\chi} \rightarrow e^{-2\,i\,\theta_M}\,\tilde{\chi} \,,
\phi_S \rightarrow e^{-i\,\theta_M}\,\phi_S \,,
\ee
with all other particles being $U(1)_M$ singlets.
In consequence, this symmetry only allows the Yukawa couplings listed in
(\ref{singlet}), (\ref{yuk1}), (\ref{yuk2}), and (\ref{majorana}).
Furthermore, there will be no coupling of the triplet $\tilde{\chi}$
to the fermion bilinear $\bar{l}_{L}\,l^{M}_{R}$ because of $U(1)_M$.
In consequence, the neutrino Dirac mass comes solely from the VEV
of $\phi_S$ as in (\ref{singlet}). 
Although this symmetry forbids the left-handed neutrino to acquire
a Majorana mass at tree level, it arises at the one-loop level as given
by
\be
\label{ml}
M_L = \lambda \, \frac{1}{16\, \pi^2}\,\frac{m_{\nu}^{D\, 2}}{M_R}\,
\ln \frac{M_R}{M_{\phi_S}}\,,
\ee
where $\lambda$ is the quartic
coupling of $\phi_S$, $M_{\phi_S}$ is the mass of $\phi_S$ 
and $m_{\nu}^D$ and $M_R$
are given by Eq.(\ref{neumass}) and Eq.(\ref{majmass}) respectively.
Notice that $M_L$ as given by (\ref{ml}) is at most two orders of magnitude
smaller than a typical see-saw light mass $m_{\nu}^{D\, 2}/M_R$
for $\lambda<1$. To be
general, we shall keep it in the mass matrix below.
Another important remark is in order here. Unlike scenarios in
which neutrino Majorana masses arise either from a singlet 
\cite{chikashige} or triplet Higgs field \cite{gelmini} 
and where there is an appearance of a massless
Nambu-Goldstone (NG) boson- the so-called Majoron- with severe
constraints, our model entails no such NG boson as we shall see
below. Also, in the aforementioned model with a Higgs triplet,
a very small VEV is given to that triplet in order to maintain
the approximate relationship $\rho =1$. For a larger VEV,
a custodial symmetry is required to guarantee $\rho =1$ at
tree-level, a topic to be discussed below.

%From (\ref{neumass}) and (\ref{majmass}), 
The Majorana mass matrix
is given by
\be
\label{majmatrix}
{\cal M} = \left( \begin{array}{cc}
M_L& m_{\nu}^D \\
m_{\nu}^D & M_R
\end{array} \right) \,,
\ee
where $m_{\nu}^D$, $M_R$ and $M_L$
are given by (\ref{neumass}), (\ref{majmass}) and (\ref{ml}) respectively.
If $g_{Sl} \sim O(g_M)$ and $v_M \gg v_S$, the eigenvalues are
approximately $M_L -(m_{\nu}^D)^2/M_R = -(g_{Sl}^{2}/g_M)\, (v_S/v_M)\,v_S
(1- \epsilon)$ (with $\epsilon < 10^{-2}$)
and $M_R$. What would be the consequences of the assumption
$v_M \sim \Lambda_{EW}$? Before discussing the constraint from the
$\rho$ parameter which would require a custodial symmetry, let us
estimate the scale $v_S$ by using the aforementioned assumption.
If $(g_{Sl}^{2}/g_M) \sim O(1)$, the constraint $m_{\nu} \leq 1\,eV$
gives 
\be
\label{v_S}
v_S \approx \sqrt{v_M \times 1\,eV} \sim O(10^{5}\,eV) \,.
\ee
In this scenario, the singlet VEV is seen to be about six orders
of magnitude smaller than the electroweak scale.

One might ask about the hierarchy between the singlet VEV and
the electroweak scale, namely the question about which mechanism that
can exist to protect the smallness of the singlet VEV. First, let us
notice that $v_S/\Lambda_{EW} \sim 10^{-6}$. This hierarchy is not 
as severe as the one that one encounters in a generic Grand Unified
Theory where at least 13 to 14 orders of magnitude difference exists
between the GUT scale and the electroweak scale. One could for instance
``fine-tune'' the cross coupling between the singlet and triplet Higgs 
fields to be less than $10^{-12}$, although this may appear ``unnatural''.
A more interesting possibility might be a scenario in which the
effective Dirac mass of the neutrino is proportional to the present value
of the singlet Higgs field $\phi_{S}(t_0) \sim 10^{5}\,eV$ whose
effective potential might be of the ``slow-rolling'' type. This
type of scenario was proposed in a mass-varying neutrino model of
the first reference of \cite{sterile}. The true minimum might be
characterized by $v_S \gg 10^{5}\,eV$. This possibility is
under investigation.

It is well-known that an introduction of Higgs representations
other than $SU(2)_L$ doublets without making sure that there
is a remaining $SU(2)$ custodial symmetry would spoil the tree-level
result $\rho =1$. One of such representations is the $SU(2)_L$ triplet
scalar which has been widely studied \cite{triplet} for various
reasons. With only one triplet, e.g. $\tilde{\chi}$, one would obtain
$\rho=2$ \cite{chanowitz}. However, it is shown in \cite{chanowitz}
that the custodial symmetry is preserved, i.e. $\rho =1$, if one has 
two triplets, one with $\tilde{\chi}(3,Y/2=1)$
and $\xi=(3,Y/2=0)$.
The two triplets, when combined, form the $(3,3)$ representation
under the global $SU(2)_L \otimes SU(2)_R$ symmetry as follows
\be
\label{chi}
\chi = \left( \begin{array}{ccc}
\chi^{0} &\xi^{+}& \chi^{++} \\
\chi^{-} &\xi^{0}&\chi^{+} \\
\chi^{--}&\xi^{-}& \chi^{0*}
\end{array} \right) \,.
\ee
The VEV of $\chi$, namely
\be
\label{chivev}
\langle \chi \rangle = \left( \begin{array}{ccc}
v_M &0&0 \\
0&v_M&0 \\
0&0&v_M
\end{array} \right) \,,
\ee
breaks the global symmetry $SU(2)_L \otimes SU(2)_R$ down to
the custodial $SU(2)$ and thus guaranteeing $\rho =1$.
As shown 
in \cite{chanowitz}, the W and Z masses have the standard expressions
$M_W = g\,v/2$ and $M_Z = M_W/\cos \theta_W$, with $v= \sqrt{v_2^2
+ 8\,v_M^2}$ and where $\langle \Phi \rangle = v_2/\sqrt{2}$
and $\langle \chi_0 \rangle = \langle \xi_0 \rangle = v_M$.
This scenario can accommodate even the case when $v_M > v_2$.
As \cite{chanowitz} already discussed, the $U(1)_M$ symmetry in
(\ref{U1M}) is broken {\em explicitly} by terms in the potential which
mix $\xi$ with $\chi$ since $\xi$ does not carry $U(1)_M$ quantum numbers.
Furthermore, such an explicit breaking term is needed in order 
to have a proper vacuum alignment. As a result, the massless NG bosons
are absent in this model. Our model also contains a singlet $\phi_S$ which
carries a  $U(1)_M$ quantum number. Unlike the ``Majoron'' case, 
our model does not generate a NG boson from the singlet since
$U(1)_M$ is already explicitly broken. The discussion of the
Higgs potential and its associated implications is beyond the scope of
this paper and will be presented elsewhere.
(As noted in
\cite{chanowitz}, the Yukawa coupling (\ref{majorana}) breaks the custodial
symmetry but its contribution to $\rho$ can be small when
$\nu_R$ and $e^{M}_R$ are near degenerate.) Let us notice
also that there is no Majorana term similar to (\ref{majorana})
involving $\xi$. The two-triplet scenario is particularly
relevant to our model because it can preserve the custodial
symmetry while allowing for $v_M$ to be of the order of the electroweak
scale. It is worth noticing at this point that the S parameter
can be made small in models with mirror fermions and more than
one Higgs doublet \cite{hung2}.
Also it is worth mentioning that that the would-be Majoron has
a mass higher than the Z boson mass and therefore does not affect
the Z width.

We conclude that the right-handed
neutrino mass $M_R$ is restricted
to a rather {\em narrow} range 
\be
\label{MRrange}
\frac{M_Z}{2} \leq M_R < 246\, GeV \,,
\ee
where the lower bound comes from the experimental Z-width requirement
and the upper bound is discussed above.
As a result, this model is
quite {\em predictive} in terms of ``detection'' of
the right-handed neutrino.

The SM
fermions and their mirror counterparts are contained the 
following $E_6$ representations: $\textbf{27}_{L}$ and
$\textbf{27}_{L}^{c}$. The details of the organization
of the SM and mirror particles in these two representations
are given in \cite{hung2}. For our purpose here, one needs just
the parts that are relevant to the above discussions.
First $l^{M}_{R}$, as defined in (\ref{rhdoublet}), 
which is equivalent to $l^{M,c}_{L}$ is
grouped in $\textbf{27}_{L}$ while the SM doublet $l_L$
is put in $\textbf{27}_{L}^{c}$. First, the neutrino Dirac mass
term as written in (\ref{singlet}) can be seen to come from
$\textbf{27}_{L}^{c,T}\sigma_{2} \textbf{27}_{L}$. Since
$\textbf{27}_{L}^{c,T}\sigma_{2} \textbf{27}_{L} \sim \textbf{1}
+ \textbf{78} + \textbf{650}$, the above fermion bilinear can 
couple to a {\em singlet} Higgs field as follows:
$\textbf{27}_{L}^{c,T}\sigma_{2} \textbf{27}_{L}\,\phi_S
(\textbf{1})$, where $\phi_S(\textbf{1})$ is an $E_6$ singlet.
The Majorana mass term coming from (\ref{majorana}) can
be seen to arise from $\textbf{27}_{L}^{T}\sigma_{2} \textbf{27}_{L}$.
Since $\textbf{27}_{L}^{T} \sigma_{2} \textbf{27}_{L} \sim
\textbf{27} + \textbf{351} +\textbf{351}^{'}$, one can see that
the Higgs representation that contains an $SU(2)_L$ triplet
which is $SU(3)_c$ and $SU(2)_Z$ singlet is 
$\phi(\textbf{351}^{'})$ with
$\textbf{351}^{'} \supset (1, \textbf{105}^{'})$, where
$\textbf{105}^{'} \supset (1,6) \supset (1,3)$ under 
$SU(3)_c \otimes SU(3)_L \otimes U(1) \rightarrow 
SU(3)_c \otimes SU(2)_L \otimes U(1)_Y$. 

The first immediate experimental implication of an electroweak
scale $M_R$ is the possibility of
directly ``detecting'' the right-handed Majorana neutrino
at colliders (Tevatron, LHC, or ILC).
One should notice: 1) $\nu_R$ 
interacts
with the W and Z bosons  since it is part of an $SU(2)_L$ doublet;
2) both $\nu_R$
and $e^{M}_{R}$ interact with $\nu_L$ and $e_L$ respectively
through the singlet scalar field $\phi_S$. In particular, since
$\phi_S$ is expected to have a mass of $O(10^{5}\,eV)$, one would have
the following interesting decay modes: $\nu_R \rightarrow \nu_L +
\phi_S$ and $e^{M}_{R} \rightarrow e_L + \phi_S$. If the mass
of $\nu_R$ is close to but less than that of $e^{M}_{R}$, one could have,
for instance, $e^{M}_{R} \rightarrow \nu_R + e_L + \bar{\nu}_L$ followed
by $\nu_R \rightarrow \nu_L + \phi_S$. Also, the heaviest $\nu_R$
could be pair-produced through $q + \bar{q} \rightarrow Z
\rightarrow \nu_R +\nu_R$ with each $\nu_R$ decaying into
a lighter $e^{M}_{R}$ plus a real or virtual $W$
followed by $e^{M}_{R} \rightarrow e_L + \phi_S$ at a ``displaced'' vertex.
Since $\nu_R$ is a Majorana particle, one could have
e.g. $e^{M,-}_{R}+ W^{+} + e^{M,-}_{R}+ W^{+} \rightarrow
e^{-}_L+e^{-}_L + W^{+} + W^{+} +2\, \phi_S$, a {\em like-sign}
dilepton event which is
distinctively different from the Dirac case. This would constitute
a high energy equivalent of the well-known neutrinoless double beta
decay. The details of the phenomenology are under investigation. 

One might ask whether or not the lightest $\nu_R$ might pose a
problem with the total energy density. Fortunately, it is unstable
because of the decay mode $\nu_R \rightarrow \nu_L + \phi_S$.
$\phi_S$, in turns, can annihilate each other when  $T < m_{\phi_S}$
as in $\phi_S + \phi_S^{*} \rightarrow \nu_L + \bar{\nu}_L$. The
remnants of these processes are the light $\nu_L$'s and the sum of
the masses of the latter is constrained by cosmology to  be less
than $1\,eV$. Furthermore, by the time when nuclesynthesis
was supposed to take place at $T \sim MeV$, the only neutrinos
that remained were the light $\nu_L$'s whose number is restricted
to be around three. The electroweak-scale right-handed neutrinos have
practically disappeared by then and therefore do not affect the big bang 
nucleosynthesis.

There is a possibility that the decays of the triplet 
Higgs particles with masses
close to the electroweak scale can wash out existing lepton asymmetry
when $T \leq M_{\chi}$.
If that happens then one might need a new mechanism for generating
the required baryon asymmetry. Whether or not this is a problem will
depend on the details of the decays of the triplet Higgs fields. This
is under investigation.

Since neutrino masses, in our scenario, come from from
an entirely different source ($SU(2)_L$ singlet and triplet Higgs fields
instead of the SM doublet), 
one might expect the leptonic ``CKM'' matrix 
to be quite different from the quark CKM matrix,
which appears to be the case experimentally. 
Since the (small) neutrino Dirac mass scale is associated with a singlet
scalar, it might be tempting to connect it with scenarios of mass-varying
neutrinos \cite{sterile}.

In summary, a model is constructed in which the Majorana mass of the
right-handed neutrinos coming from the VEVs of
$SU(2)_L$-triplet Higgs fields  is found to obey $\frac{M_Z}{2} \leq M_R < 246\, GeV$,
without violating the constraint $\rho=1$ at tree level because
of the presence of a custodial symmetry. Its interest lies in the possibility of
producing and detecting the right-handed neutrinos at current and future colliders, 
a prospect that is not present with a generic see-saw scenario where $M_R$ is typically
of the order of the GUT scale.

\begin{acknowledgments}
I would like to thank Goran Senjanovic, Vernon Barger
and Paul Frampton for discussions.
This work is supported
in parts by the US Department of Energy under grant No.
DE-A505-89ER40518.
\end{acknowledgments}

\end{document}